\documentclass[twocolumn]{aastex701}

\usepackage{amsmath,amssymb}
\usepackage{float}
\usepackage{hyperref}
\usepackage{dsfont}
\usepackage{multirow}

\begin{document}

%\title{A PANORAMIC View of RUBIES through Time: \\ Connecting Times Since Quenching to Quiescent Galaxy Number Densities up to $z\sim7$}

\title{Winding Back the Clock: Recent Star Formation Histories of Massive Quiescent Galaxies Are Consistent With Their Rapid Number Density Evolution Since $\mathbf{z\sim7}$}

\shortauthors{Zhang et al.}

\author[0000-0001-6454-1699]{Yunchong Zhang} 
\affiliation{Department of Physics and Astronomy and PITT PACC, University of Pittsburgh, Pittsburgh, PA 15260, USA}
\email[show]{yuz369@pitt.edu}

\author[0000-0001-7673-2257]{Zhiyuan Ji} 
\affiliation{Steward Observatory, University of Arizona, 933 N. Cherry Avenue, Tucson, AZ 85721, USA}
\email{zhiyuanji@arizona.edu}

\author[0000-0001-5063-8254]{Rachel Bezanson}
\affiliation{Department of Physics and Astronomy and PITT PACC, University of Pittsburgh, Pittsburgh, PA 15260, USA}
\email{rachel.bezanson@pitt.edu}

\author[0000-0003-2919-7495]{Christina C. Williams}
\affiliation{NSF National Optical-Infrared Astronomy Research Laboratory, 950 North Cherry Avenue, Tucson, AZ 85719, USA}
\affiliation{Steward Observatory, University of Arizona, 933 N. Cherry Avenue, Tucson, AZ 85721, USA}
\email{christina.williams@noirlab.edu}

\author[0000-0003-2680-005X]{Gabriel Brammer}
\affiliation{Cosmic Dawn Center (DAWN), Copenhagen, Denmark} 
\affiliation{Niels Bohr Institute, University of Copenhagen, Jagtvej 128, K{\o}benhavn N, DK-2200, Denmark}
\email{}

\author[0000-0001-9978-2601]{Aidan P. Cloonan}\thanks{NSF Graduate Research Fellow}
\affiliation{Department of Astronomy, University of Massachusetts, Amherst, MA 01003, USA}
\email{}

\author[0000-0002-2380-9801]{Anna de Graaff}
\thanks{Clay Fellow}
\affiliation{Center for Astrophysics, Harvard \& Smithsonian, 60 Garden St, Cambridge, MA 02138, USA}
\affiliation{Max-Planck-Institut f\"ur Astronomie, K\"onigstuhl 17, D-69117 Heidelberg, Germany}
\email{anna.de_graaff@cfa.harvard.edu}

\author[0000-0002-5612-3427]{Jenny E. Greene}
\affiliation{Department of Astrophysical Sciences, Princeton University, 4 Ivy Lane, Princeton, NJ 08544, USA}
\email{jgreene@astro.princeton.edu}

\author[0000-0002-3301-3321]{Michaela Hirschmann}
\affiliation{Institute of Physics, Laboratory for Galaxy Evolution, EPFL, Observatory of Sauverny, Chemin Pegasi 51, CH-1290 Versoix, Switzerland}
\email{michaela.hirschmann@epfl.ch}

\author[0000-0002-8896-6496]{Christian Kragh Jespersen}
\affiliation{Department of Astrophysical Sciences, Princeton University, 4 Ivy Lane, Princeton, NJ 08544, USA}
\email{cj1223@princeton.edu}

\author[0000-0002-3475-7648]{Gourav Khullar}
\affiliation{Department of Astronomy, University of Washington, Physics-Astronomy Building, Box 351580, Seattle, WA 98195-1700, USA}
\affiliation{eScience Institute, University of Washington, Physics-Astronomy Building, Box 351580, Seattle, WA 98195-1700, USA}
\email{gkhullar@uw.edu}

\author[0000-0003-3021-8564]{Claudia del P. Lagos}
\affiliation{International Centre for Radio Astronomy Research (ICRAR), M468, University of Western Australia, 35 Stirling Hwy, Crawley, WA 6009, Australia}
\email{claudia.lagos@uwa.edu.au}

\author[0000-0001-6755-1315]{Joel Leja}
\affiliation{Department of Astronomy \& Astrophysics, The Pennsylvania State University, University Park, PA 16802, USA} 
\affiliation{Institute for Computational \& Data Sciences, The Pennsylvania State University, University Park, PA 16802, USA}
\affiliation{Institute for Gravitation and the Cosmos, The Pennsylvania State University, University Park, PA 16802, USA}
\email{joel.leja@psu.edu}

\author[0000-0003-0695-4414]{Michael V.\ Maseda}
\affiliation{Department of Astronomy, University of Wisconsin-Madison, 475 N. Charter St., Madison, WI 53706 USA}
\email{}

\author[0000-0002-2446-8770]{Ian McConachie}
\affiliation{Department of Astronomy, University of Wisconsin-Madison, 475 N. Charter St., Madison, WI 53706 USA}
\email{ian.mcconachie@wisc.edu}

\author[0000-0001-5851-6649]{Pascal A. Oesch}
\affiliation{Department of Astronomy, University of Geneva, Chemin Pegasi 51, 1290 Versoix, Switzerland}
\affiliation{Cosmic Dawn Center (DAWN), Copenhagen, Denmark} 
\affiliation{Niels Bohr Institute, University of Copenhagen, Jagtvej 128, K{\o}benhavn N, DK-2200, Denmark}
\email{}

\author[0000-0002-0108-4176]{Sedona H. Price}
\affiliation{Space Telescope Science Institute, 3700 San Martin Drive, Baltimore, Maryland 21218, USA}
\email{sprice@stsci.edu}

\author[0000-0003-4075-7393]{David J. Setton} 
\thanks{Brinson Prize Fellow} \affiliation{Department of Astrophysical Sciences, Princeton University, 4 Ivy Lane, Princeton, NJ 08544, USA}
\email{davidsetton@princeton.edu}

\author[0000-0002-1714-1905]{Katherine A. Suess}
\affiliation{Department for Astrophysical \& Planetary Science, University of Colorado, Boulder, CO 80309, USA}
\email{Wren.Suess@colorado.edu}

\author[0000-0001-7160-3632]{Katherine E. Whitaker}
\affiliation{Department of Astronomy, University of Massachusetts, Amherst, MA 01003, USA} 
\affiliation{Cosmic Dawn Center (DAWN), Copenhagen, Denmark}
\email{kwhitaker@astro.umass.edu}

\begin{abstract}
Massive quiescent galaxies have been identified out to $z\sim7$ in early JWST data in a substantial excess ($\rm \gtrsim 1\,dex$ at $z>4$) of number densities from most theoretical predictions. We investigate whether the number densities implied by the star formation histories of quiescent galaxies at $2<z<5$ are consistent with the observed number density evolution of that population since $z>7$. For this work, we rely on stellar population synthesis modeling of JWST NIRCam photometry (from CEERS and PRIMER) and NIRSpec/PRISM spectra of massive ($\rm M_{*} > 10^{10.5}M_{\odot}$) quiescent galaxies in the RUBIES survey. We infer their star-formation histories through Bayesian spectro-photometric fitting with Prospector, exploring the sensitivity of our results to stellar libraries and SFH priors. For each source, we compute a timescale over which it would be identified as quiescent -- leveraging the recent and most robust SFH timescale -- and deduce the number density of the quiescent population at previous epochs. These reconstructed number densities are then compared to existing observational constraints, including a new measurement from the PANORAMIC pure parallel survey, whose wide-area and independent sightlines reduce sensitivity to cosmic variance. We find striking agreement between reconstructed and observed number densities up to $z\sim7$, a self-consistency that lends credence to stellar population synthesis modeling of distant quiescent galaxies. Furthermore, by connecting the recent ($\rm \sim 1\,Gyr$) star-formation histories and number densities of quiescent galaxies and their implied progenitors, we reinforce the known tension between observations and model predictions at $3<z<7$.
\end{abstract}


\section{Introduction} 

One of the most striking early JWST discoveries is the confirmation of a population of massive quiescent galaxies that have rapidly formed over $\rm 10^{10} M_{\odot}$ in stellar mass and subsequently quenched within merely $\rm 1 \, Gyr$ after the Big Bang \citep[e.g.,][]{Carnall.etal.2023,Valentino.etal.2023,Kakimoto.etal.2024,Antwi-Danso.etal.2025,Baker.etal.2025b,deGraaff.etal.2025a}. Early photometric JWST samples \citep[e.g.,][]{Carnall.etal.2023,Valentino.etal.2023,Long.etal.2024}, supported by smaller spectroscopic samples \citep{Nanayakkara.etal.2025,Baker.etal.2025a,Zhang.etal.2026}, indicate that the quiescent galaxies at $z\sim4$ are almost an order of magnitude more common than the predictions from most galaxy evolution simulations \citep[see][for a detailed comparison]{Lagos.etal.2025}. Relieving this tension requires models to incorporate more efficient stellar assembly followed by more rapid suppression \citep{ChandroGomez.etal.2025,Chaikin.etal.2025,Chaikin.etal.2026}. In some of the most extreme cases, the implied star-formation histories (SFHs) suggest that some massive galaxies quenched at $z>6$ \citep[e.g.,][]{Glazebrook.etal.2024,Carnall.etal.2024,McConachie.etal.2025a}. These extremely early formation solutions, which push against the limits set by our galaxy formation models, are often referred to as ``maximally old". This discovery has sparked heated debates within the community regarding the potential formation channels for these galaxies \citep[e.g.,][]{Liu.etal.2022,Dekel.etal.2023,Ferrara.etal.2023} and whether they even challenge the $\mathrm{\Lambda CDM}$ cosmology model \citep[e.g.,][]{BoylanKolchin.etal.2023}.

In the meantime, these extreme SFH solutions may not be reliable inferences, plagued by familiar and new modeling challenges. One key challenge is the well-known degeneracies between metallicity, and perhaps more importantly at these redshifts, chemical enrichment histories, and ages. Specifically, the rapid stellar assembly of these early galaxies implies that the stars will be alpha-enhanced, limiting the utility of stellar population synthesis modeling with scaled-solar metallicities \citep{Beverage.etal.2025,Park.etal.2025,Hamadouche.etal.2026}. Even at fixed abundance patterns, multi-modal SFH solutions associated with different metallicity assumptions can mean the difference between maximally old and less extreme star formation histories \citep{deGraaff.etal.2025a}.

Further doubts are raised when reconstructing the observability of these galaxies based on their SFHs (i.e., ``SFH archaeology" \footnote{Not to be confused with ``Galactic archaeology" in the local Universe, which makes use of resolved stars.}; \citealp{McConachie.etal.2025a}). We should expect to identify massive quiescent galaxies as soon as they form ($z\sim7$). However, only a few quiescent galaxies have been spectroscopically confirmed at $z>5$ to date \citep{Onoue.etal.2024,Weibel.etal.2025,Baker.etal.2026}, and none as large as the most massive ($\rm 10^{11}M_{\odot}$) ``maximally old" quiescent galaxies. Archaeological analysis of the SFHs of quiescent galaxies at $z\sim2$ implies even higher number densities than observed at $z\sim5$ \citep{Park.etal.2024}. However, interpreting this tension is complicated by the fact that SFH history reconstruction is most robust for the most recent timescales; beyond $\rm \sim1\,Gyr$, this inference becomes increasingly prior-dominated \citep{Carnall.etal.2019,Leja.etal.2019}. Therefore, interpreting the population evolution from SFHs is challenging from a model perspective and is further convoluted by the biased spectroscopic targeting in any given sample.

The massive quiescent galaxy sample at $2<z<5$ assembled in \cite{Zhang.etal.2026} enables a higher-redshift excavation site for SFH archaeology. These galaxies are observed with NIRSpec \textsc{PRISM} spectra from the JWST RUBIES (Red Unknowns: Bright Infrared Extragalactic Survey; GO\#4233, PIs: A. de Graaff and G. Brammer; \citealp{deGraaff.etal.2025b}) Program. The spectroscopic data provide for robust SFH inference, the well-defined targeting strategy reduces biases in population estimates, and at these high redshifts, recent $\sim1$ Gyr timescales account for a higher fraction of the SFHs. In this letter, we investigate whether the number densities reconstructed from the recent SFHs of quiescent galaxies in RUBIES are consistent with direct measurements at earlier epochs. Because this relatively small sample ($\rm \sim300 \,arcmin^2$) is fundamentally limited by cosmic variance, we complement this internal comparison with an additional comparison to number densities derived from the wide-area PANORAMIC survey \citep{Williams.etal.2025,Ji.etal.2026}.

The structure of this letter is as follows. In Section \ref{sec: data}, we introduce the selection of the RUBIES massive quiescent galaxy sample and associated spectroscopy and photometry. We detail the spectro-photometric modeling and SFH reconstruction, including an exploration of systematic modeling uncertainties in Section \ref{sec: Methods}. We describe our methodology for recovering the number density in previous epochs from time since quenching in Section \ref{sec: T2N}. In Section \ref{sec: results}, we present the reconstructed number density of massive quiescent galaxies using our sample, which we compare to direct observations in the literature. We discuss the implications of our results and summarize our findings in Section \ref{sec: discussion}. Throughout this paper, we assume a flat $\mathrm{\Lambda CDM}$ cosmology with $ \mathrm{\Omega_{\Lambda} = 0.71}$, $ \mathrm{\Omega_{m} = 0.29}$, and $\mathrm{H_{0} = 69.32 \, km\,s^{-1}\,Mpc^{-1}}$ from the 9-year results of the WMAP mission \citep{Hinshaw.etal.2013} and adopt a Chabrier IMF \citep{Chabrier.etal.2003}.

\section{The RUBIES Massive Quiescent Galaxy Sample} \label{sec: data}

In this work, we study the 17 massive ($\rm M_{*} > 10^{10.5} M_{\odot}$) quiescent galaxies at $2<z<5$ selected from the RUBIES program \citep{deGraaff.etal.2025b}. The selection method is presented in detail in \cite{Zhang.etal.2026}, but in brief: the sample is first narrowed down by performing principal component analysis on all RUBIES spectra at $2<z<5$, and then finalized based on specific star formation rate (sSFR) obtained from stellar population synthesis modeling. The sample includes 14 galaxies robustly identified as quiescent with $50\%$ percentile posterior sSFR less than $\rm 10^{-10} \,yr^{-1}$, as well as 3 galaxies that are identified as ``marginally quiescent" with $16\%$ percentile posterior sSFR less than $\rm 10^{-10} \,yr^{-1}$. In this work, we adopt the quiescence definition of $\rm sSFR<10^{-10} yr^{-1}$ to be aligned with the selection of this sample \citep{Zhang.etal.2026} and is commonly adopted in the literature \citep[e.g.,][]{Lagos.etal.2025,Nanayakkara.etal.2025}. Alternatively, one can adopt a time-evolving threshold such as $\rm sSFR<\frac{0.2}{t_{H}(z)}$, where $\rm t_{H}(z)$ is the age of the universe at redshift $z$ \citep[e.g.,][]{Baker.etal.2025a,Stevenson.etal.2026}. This alternative selection criterion would identify the same sample of quiescent galaxies in RUBIES, including one more galaxy at $z>4$ that is otherwise identified as ``marginally quiescent".

These galaxies have associated NIRCam imaging in F090W, F115W, F150W, F200W, F277W, F356W, F410M, and F444W from the CEERS program in the EGS field (ERS-1345, PI Finkelstein; \citealp{Finkelstein.etal.2023}) and from the PRIMER program in the UDS field (GO-1837, PI Dunlop; \citealp{Donnan.etal.2024}). Specifically, the photometry used in this analysis is extracted through customized apertures matched to the NIRSpec Micro Shutter Array (MSA) apertures, from mosaic images that are point-spread-function(PSF)-homogenized to F444W \citep{Zhang.etal.2026}. This choice of photometry is motivated by the fact that massive quiescent galaxies at these redshifts display diverse color gradients \citep{Siegel.etal.2025,Kawinwanichakij.etal.2026}, and their spatially complex color information is not fully captured by the small spectral aperture. We choose to focus on interpreting the central average color information within the spectral aperture, forgoing a comprehensive characterization of global average properties. The images are reduced with \texttt{grizli} \citep{Brammer.etal.2023a}, corresponding to version 7.2 on DJA \citep{Valentino.etal.2023}. The exact methodology for the reduction and PSF-homogenization of these images is presented in \cite{Weaver.etal.2024}. For galaxies in EGS, the photometry in F090W is not included as it is not available as part of the release by \cite{Weaver.etal.2024}. To account for the aperture loss in our customized aperture photometry, we derive a scaling correction factor for each galaxy \citep{Zhang.etal.2026}, using the ``corrected-to-total'' aperture photometry value in F444W from existing catalogs in EGS and UDS \citep{Wright.etal.2024,Cutler.etal.2024}.

The RUBIES NIRSpec \textsc{PRISM} spectra analyzed in this work were reduced with  \texttt{msaexp}\footnote{\href{https://github.com/gbrammer/msaexp}{https://github.com/gbrammer/msaexp}} \citep{Brammer.etal.2023b}, corresponding to version 3 of NIRSpec data released on DJA\footnote{\href{https://s3.amazonaws.com/msaexp-nirspec/extractions/nirspec_rubies_graded_v3.html}{https://s3.amazonaws.com/msaexp-nirspec/extractions/nirspec\_rubies\_graded\_v3.html}}. The comprehensive reduction procedure for these spectra, as well as the spectroscopic targeting strategy in RUBIES, can be found in \cite{deGraaff.etal.2025b,Heintz.etal.2025}. The original observations associated with the data product analyzed in this work can be accessed via \dataset[doi: 10.17909/sjsj-8p46]{https://doi.org/10.17909/sjsj-8p46}.

\section{Reconstructing the Star Formation Histories} \label{sec: Methods}
\subsection{Prospector modeling} 
In order to reconstruct the number density of quiescent galaxies prior to observation, we estimate the lookback time period during which the given galaxy would have been selected as quiescent. Throughout this letter, we approximate this timescale as a modified time since quenching $t_{q}' \equiv t_{q} - \rm 50 \,Myrs$, where the $t_{q}$ is the time since the moment when the sSFR drops below $\rm 10^{-10} yr^{-1}$. We subtract $\rm 50 \,Myrs$ from $t_{q}$ to account for the period during which the massive O and B-type stars remain visible\footnote{Subtracting $\rm 50 \,Myrs$ slightly suppresses the quiescent visibility and as well as the reconstructed number density, although it doesn't change our qualitative conclusions.}. 

To first estimate time since quenching for each galaxy, we jointly model the NIRSpec \textsc{PRISM} spectrum and NIRCam photometry for each galaxy, using the Bayesian stellar population inference code \texttt{Prospector} \citep{Leja.etal.2017,Johnson.Leja.2017,Johnson.2021}. The posterior sampling is performed with the nested sampling code \texttt{dynesty} \citep{Speagle.etal.2020}. \texttt{Prospector} makes use of the stellar population synthesis (SPS) models from the Flexible Stellar Population Synthesis (FSPS) package \citep{Conroy.etal.2009,Conroy.Gunn.2010}. We use a polynomial of order 5 to flux calibrate the spectrum to photometry. We enforce a minimum uncertainty floor of 5\% on both the spectrum and photometry in this analysis in order to avoid overconfident posteriors from model mis-specification \citep[e.g., ][]{Muzzin.etal.2013,Jespersen.etal.2025a}. Since the input photometry is measured from a small aperture, we account for the aperture loss by applying a scaling factor to any relevant model outputs, such as stellar mass and SFH. For each galaxy, we take the scaling factor as the ratio between the F444W fluxes measured in the slit-like aperture and the catalog F444W fluxes that are ``corrected-to-total" \citep{Weaver.etal.2024}, following \cite{Zhang.etal.2026}. The mean scaling factor is $\sim 4$ in this sample.

To probe the systematic uncertainties in SFH reconstruction due to model setup choices, we fit each galaxy with three different variants: the fiducial model, the bursty SFH prior variant, and the C3K spectral library\footnote{Descriptions on the C3K spectral library can be found in \cite{Park.etal.2025}.} variant. In the fiducial model, we adopt the MILES spectral library \citep{Sanchez-Blazquez.etal.2006,Falcon-Barroso.etal.2011} and a non-parametric SFH that utilizes the \texttt{Prospector} continuity prior described in \cite{Leja.etal.2019}. In the bursty variant, we keep using the MILES spectral library and adopt a bursty prior \citep{Tacchella.etal.2022} for the star formation rate ratios between different age bins in the non-parametric SFH. The bursty prior does not necessarily prefer a bursty SFH solution over a continuous solution, but instead reduces the penalty on more abrupt changes in the SFH. Finally, in the C3K variant, we keep a continuity prior for the SFH and adopt the C3K spectral library.

We adopt a redshift-dependent age binning scheme as follows. For the most recent $\mathrm{200\,Myr}$ in lookback time, we place four logarithmically-spaced age bins with widths of $\mathrm{10,40,50}$,and ${\rm 100\,Myr}$. We then linearly add four bins of $\mathrm{200\,Myr}$ until we reach $\mathrm{ 1\,Gyr}$ in lookback time. The remaining time is evenly divided into $N_\mathrm{old}$ bins. We calculate $N_\mathrm{old}$ by taking the ceiling of $(t_\mathrm{universe}-1\,\mathrm{Gyr})/0.5\,\mathrm{Gyr}$, where $t_\mathrm{universe}$ is the age of the universe, resulting in 1 to 5 old bins for our sample at $2<z<5$. The specific choice of these age bins is to have a consistent resolution of time since quenching in the most recent $\mathrm{ 1\,Gyr}$ for all objects. Only four objects in our sample are identified as ``old quiescent", which could have quenched over $\mathrm{ 1\,Gyr}$ ago. However, constraining the time since quenching beyond $\mathrm{ 1\,Gyr}$ is intrinsically uncertain, as the stellar differentiability (i.e. difference in simple stellar populations) scales logarithmically with age. For these galaxies, we expect the broad old age bins will encapsulate the large uncertainty in their times since quenching.

\begin{figure*}[ht!]
\centering
\includegraphics[width = 1\textwidth]{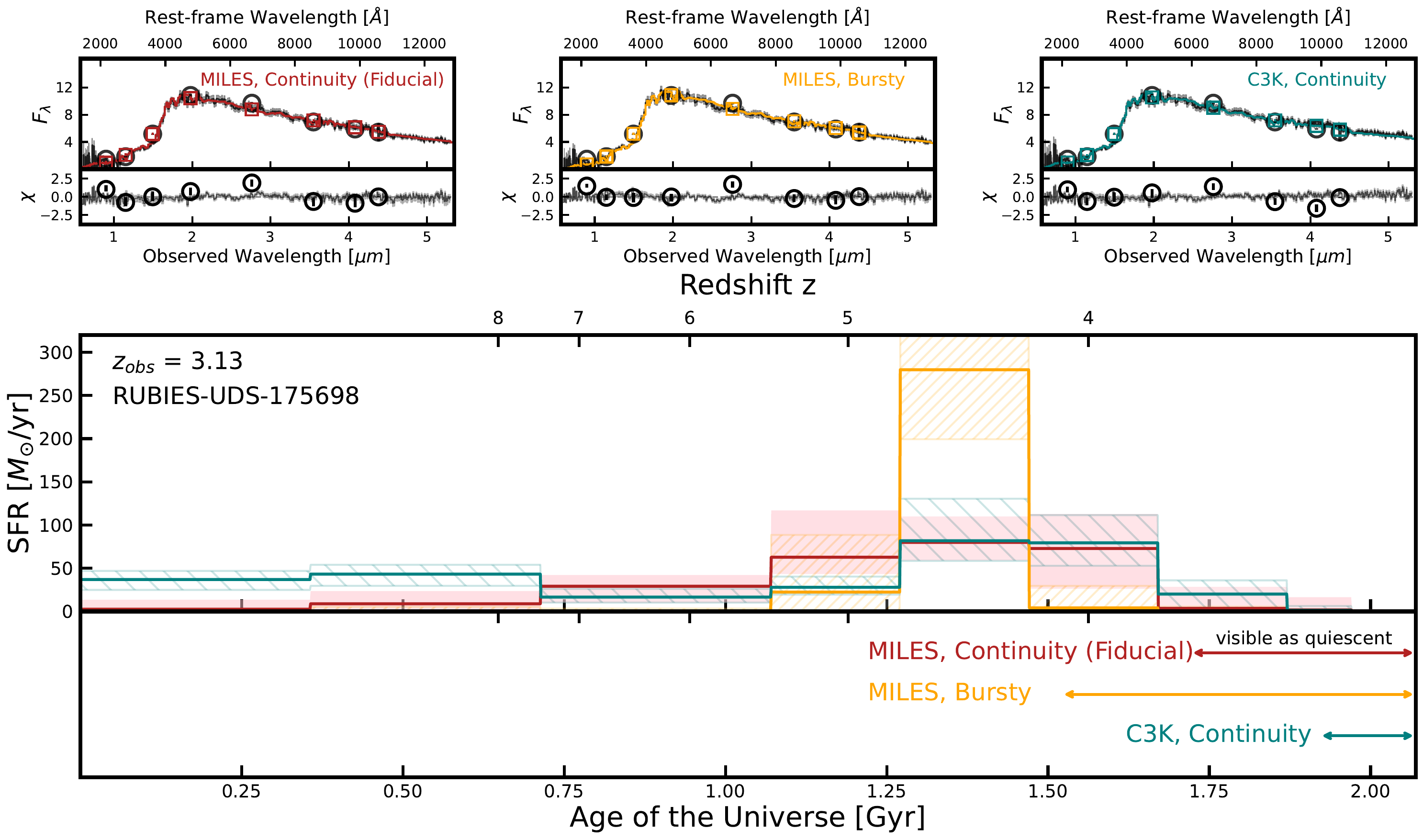}
\caption{Top panels: The spectro-photometric fits of an example quiescent galaxy (RUBIES-UDS-175698) at $z\sim3.1$, given three \texttt{prospector} setups. The fiducial model (red) adopts the MILES spectral library and a continuity SFH prior. Additionally, we test a second model variant that changes to a bursty SFH prior (orange) and a third model that changes to the C3K spectral library (teal). The model spectrum and photometry are shown in color, and the observed spectrum and photometry are shown in black and gray. Middle panel: The inferred SFHs of this galaxy. For each setup, we show the resulting median (solid lines) and $16-84\%$ percentile (solid or hatched bands) SFHs, adopting the same color coding as the top panels. Bottom panel: The median time since quenching in each model, defined as the lookback time since $\rm sSFR$ drops below $\rm 10^{-10} yr^{-1}$, with $\rm 50 \,Myrs$ subtracted to account for O and B star lifetimes. For a given quiescent galaxy, this timescale is sensitive to modeling assumptions, even though the model fits to the observed data are similarly excellent. 
\label{fig: method}}
\end{figure*}

For other modeling details, we closely follow \cite{deGraaff.etal.2025a,Zhang.etal.2026}. We adopt MIST isochrones \citep{Choi.etal.2016,Dotter2016} and assume a Chabrier IMF \citep{Chabrier.etal.2003}. We fix the redshift to \texttt{msaexp}-derived spectroscopic redshifts. We assume a two-parameter \cite{Kriek.Conroy.2013} dust law, which is parameterized by the attenuation around old ($\mathrm{t > 10 \,Myr}$) stars fit in the range $\tau \in [0,2.5]$ and a free dust index $\delta \in [-1,0.4]$ that describes the deviation from the \cite{Calzetti.etal.2000} dust law and includes a UV bump that depends on the slope parameterized as in \cite{Noll.etal.2009}. The attenuation around young ($\mathrm{t < 10 \,Myr}$) stars is fixed to be twice that of the older populations. The stellar metallicity is set as free, with a logarithmically sampled uniform prior in the range $\mathrm{log(Z/Z_{\odot}) \in \,[-2, 0.2]}$, marginalizing over a wide range, since robust constraints on stellar metallicity of early quiescent galaxies remain controversial. We also note that the stellar metallicity is set as a constant (non-time-evolving) in these models, although local universe studies have found that an evolving metallicity assumption is required to better recover SFHs \citep{Bellstedt.etal.2020}.

For galaxies at $\rm z \gtrsim 4$, we mask all wavelengths shorter than rest-frame $\mathrm{1200 \AA}$ to avoid contributions from intergalactic medium absorption. Some of the galaxies in this sample have emission lines that can be due to AGN activity and are complicated to model and disentangle from star formation. We opt to marginalize over all emission lines by fitting Gaussian profiles. The specific list of lines and Gaussian profile fitting strategy is identical to \cite{Zhang.etal.2026}. Before fitting, all model spectra are convolved with a line spread function that is a factor of 1.3 narrower than the original JWST User Documentation (JDox) curves to account for instrumental dispersion\footnote{This convolution is only applied to the wavelength regime where the library spectral resolution is much higher than the resolution of the observed spectrum, which is $\rm 3525-7500\AA$ for MILES and $\rm 2750-9100\AA$ for C3K.}, following \cite{deGraaff.etal.2025a}. This is motivated by the fact that the line spread function curves in JDox are broader than those measured in practice \citep{deGraaff.etal.2024}. We also include two free velocity dispersions that smooth the stellar continuum and ionized gas emission, which we allow to vary in the range $\mathrm{[0,1000] \,km/s}$ to marginalize over the uncertainty in the line spread function and the intrinsic dispersion of the galaxy.

In figure \ref{fig: method}, we showcase the reconstruction of SFH and estimation of quiescent visibility timescales for an example galaxy at $z\sim 3.1$. While all three models provide decent fits to the observed spectrum and photometry, each leads to diverse SFHs and implies widely different $t_{q}$ (and consequently $t_{q}'$). Systematic differences in $t_{q}$ due to these modeling choices can impact our interpretation of the observability of these galaxies as quiescent in the past. We compare the systematic differences in these $t_{q}$ recovered by different model setups in the next Section. We detail our method of converting these timescales to number densities in prior epochs in Section \ref{sec: T2N}. 

\subsection{Systematic differences in inferring time since quenching: priors and spectral libraries} 

\begin{figure*}[ht!]
\centering
\includegraphics[width = 1\textwidth]{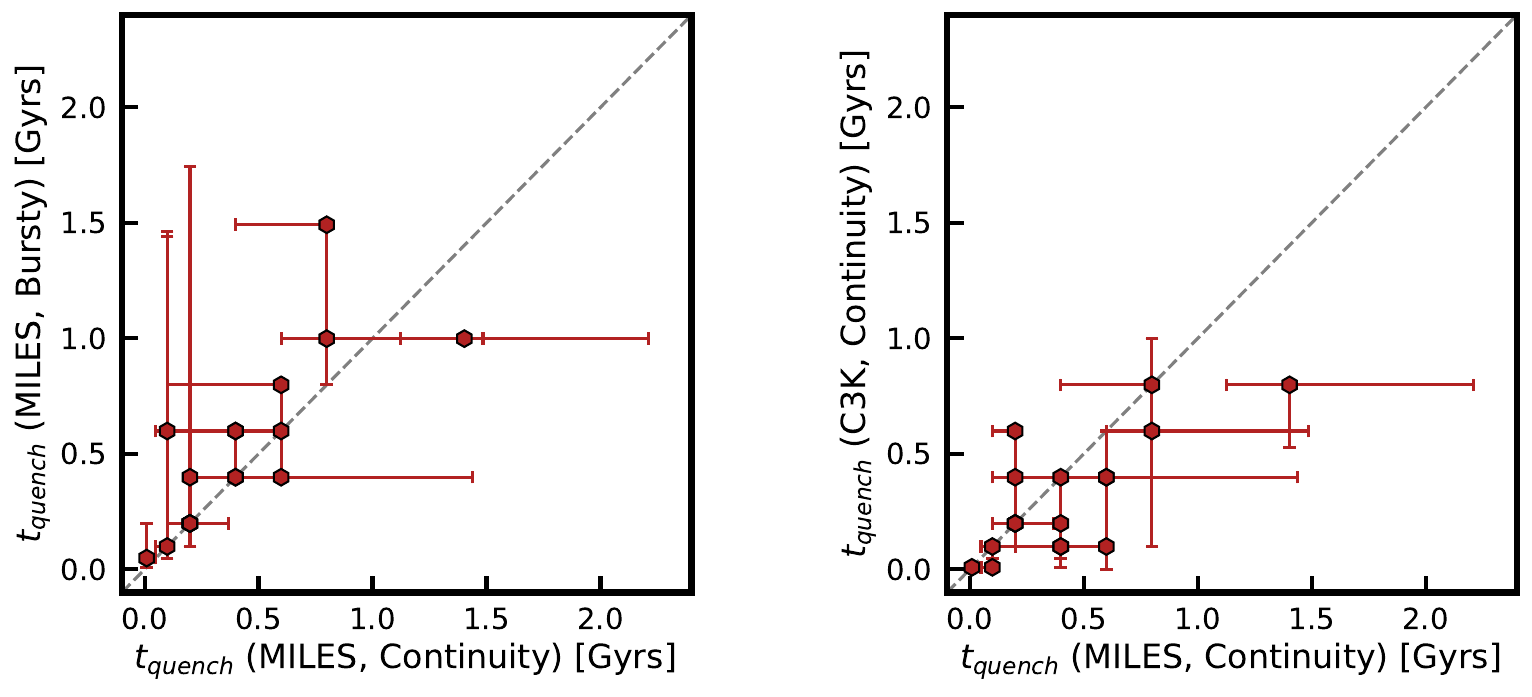}
\caption{Recovered $t_{q}$ between given pairs of model setups. In the left panel, we show the fiducial model (MILES spectral library; continuity SFH prior) versus the bursty prior variant. In the right panel, we compare the fiducial model to the variant adopting the C3K spectral library. Overall, we find that models with a bursty SFH prior systematically infer longer time since quenching than those with a continuity prior. In addition, models with the C3K spectral library systematically infer shorter $t_{q}$ than those adopting MILES.
\label{fig: timescale}}
\end{figure*}

In figure \ref{fig: timescale}, we compare $t_{q}$ (time since sSFR drops below $\rm 10^{-10} yr^{-1}$) inferred from different model variants for each galaxy. We show the fiducial model versus the bursty variant and the C3K variant in the left and right panels, respectively. We only show galaxies in each panel if $50\%$ of the sSFR posterior in the most recent bin is below the $\rm 10^{-10} yr^{-1}$ threshold in each variant. Overall, we find the bursty model variant systematically predicts earlier $t_{q}$ than the fiducial model, which corresponds to a systematically higher probability of catching these galaxies as quiescent in the past. We also find the C3K model variant systematically predicts longer $t_{q}$ than the fiducial model. This systematic uncertainty due to modeling assumptions translates to an even larger variation in number densities than exhibited by the range of theoretical predictions \citep{Lagos.etal.2025}, emphasizing the importance of such tests.

The bursty SFH prior permits more abrupt changes in SFR between neighboring bins, thus SFR can decline rapidly and yield longer $t_{q}$. A similar trend is also reported in \cite{Park.etal.2024}. The origin of the systematic shifts introduced by the C3K spectral library is less clear. Previous studies of some high-redshift recently quenched galaxies have found multi-modal SFH solutions, which are typically correlated with different metallicity solutions \citep{deGraaff.etal.2025a}. Constraining the stellar metallicity is challenging at NIRSpec \textsc{PRISM} resolution ($R\sim 100$); therefore, the resulting solutions are highly degenerate. In this sample, we do find that galaxy fits with shorter $t_{q}$ from the C3K library versus MILES tend to prefer higher stellar metallicity. It is likely that \texttt{prospector} preferentially samples one of the modes when adopting the C3K spectral library and the other with MILES. Fully resolving this issue likely requires additional data and is beyond the scope of this work. 

Since non-parametric $\rm SFR(t)$ only varies at the edges of time bins, the best-fitting $t_{q}$ and associated uncertainties are artificially discretized. This may truncate the $t_{q}$ posteriors and thus not encapsulate the difference between true time since quenching and the inferred $t_{q}$. Furthermore, the SFR is modeled as an averaged value in each time bin, which could further bias timescales. In the age bin where quenching occurs, the high previous SFR could elevate the averaged SFR and fall below an sSFR threshold at a later bin edge, systematically underestimating $t_{q}$. However, the amplitude of these systematics should always be smaller than the width of the nearest time bin and therefore be subdominant to the modeling uncertainties. Incorporating more flexibility in bin edges \citep[e.g.,][]{Suess.etal.2022} can mitigate this effect and more precisely constrain timescales; we defer this exploration to future studies. A different definition of quiescence (i.e., a constant sSFR criterion versus a time-evolving sSFR criterion) could impact these timescales, but given the dramatic drops in sSFR at bin edges, we expect this difference to be negligible relative to the uncertainty introduced by discretized age bins. Finally, future efforts should also examine whether different SED fitting codes \citep[e.g., BEAGLE, CIGALE][]{Chevallard.etal.2016,Boquien.etal.2019} or more complicated stellar population models \citep[e.g., accounting for binary evolution or stellar rotation][]{Eldridge.etal.2017,Dorn-Wallenstein.etal.2020} lead to any systematic effects on $t_{q}$.

\section{From time since quenching to number density}\label{sec: T2N}

With modified time since quenching, $t_{q}'$, for each target (the lookback time since $\rm sSFR<10^{-10}yr^{-1}$ minus $\rm 50\,Myrs$ to account for the remaining visibility of O/B-type stars), we evaluate the probability that each would have been selected as quiescent in previous epochs as follows. For a quiescent galaxy and any given redshift bin [$z_{i}$,$z_{i+1}$] prior to their observed redshift, we define a visibility factor:
\begin{equation}\label{eq: visibility}
\mathcal{P}(\mathrm{t_{q}'}) =
\begin{cases}
1 & \text{if } t_{q}' > t(z_{i+1}), \\
\frac{(t_{q}'-t(z_{i}))}{(t(z_{i+1})-t(z_{i}))} & \text{if } t(z_{i})<t_{q}' < t(z_{i+1}),\\
0 & \text{if } t_{q}'<t(z_{i})\text{,}
\end{cases}
\end{equation}
where $ t(z_{i})$ denotes the lookback time at redshift $z_{i}$. The visibility factor serves as a simple probabilistic correction to the effective number count of a quiescent galaxy, which equals the fraction of the time available in the redshift bin that it would appear as quiescent. 

The target selection in RUBIES was designed to prioritize bright red sources with a selection function that has a completeness inversely proportional to the space density of sources in F444W -- photo-z -- F150W-F444W-color space \citep{deGraaff.etal.2025b}. For massive quiescent galaxies, one can approximately correct for the selection effect by applying a multiplicative factor to the effective number count, following \cite{Zhang.etal.2026}. This factor is essentially the inverse of survey selection completeness in a given color-magnitude-redshift bin ($\mathbf{m}$) that contains the observed quiescent galaxy: 
\begin{equation}
\mathcal{S}_{tot}(\mathbf{m})^{-1} = N_{total,\mathbf{m}}/N_{surveyed,\mathbf{m}},
\end{equation}
where $N_{total,\mathbf{m}}$ is the total number of available objects in the NIRCam footprints of UDS and EGS and $N_{surveyed,\mathbf{m}}$ is the total number of objects surveyed with a robust quality spectrum. 

For a given population of N quiescent galaxies observed in an epoch [$z_{j}$,$z_{j+1}$], we then recover the progenitor population number density [$z_{i}$,$z_{i+1}$] as follows:
\begin{equation} \label{eq: number density}
\mathrm{n_{Massive,Q}}  =\frac{1}{\mathrm{V_{eff}}}\sum^{N} 1 \cdot \mathcal{S}_{tot}^{-1} \cdot \mathcal{P},
\end{equation}
where $\rm n_{Massive,Q}$ is the number density of massive quiescent galaxies.
The effective volume $\mathrm{V_{eff}}$ is the co-moving volume set by the survey angular area in the observed epoch, which is given by:
\begin{equation}
 \mathrm{V_{eff} = \frac{\Omega_{field}}{3}\cdot[d_{com}(z_{j+1})^{3}-d_{com}(z_{j})^{3}]},
\end{equation}
where $\rm \Omega_{field}$ is the combined angular size of the survey fields (EGS and UDS; $\rm \sim 300 \,arcmin^2$) and $\rm d_{com}(z)$ is the co-moving distance to redshift $\rm z$. In this work, we perform these calculations for quiescent galaxies observed in redshift bins of $[2,3]$,$[3,4]$, and $[4,5]$, respectively, and reconstruct their corresponding number densities in redshift bins of $[3,4]$, $[4,5]$, and $[5,7]$.

Our models do not produce any rejuvenation solutions except for one galaxy (RUBIES-UDS-121002, ``marginally quiescent"; \citealp{Zhang.etal.2026}) at $z\sim 2.6$. Nevertheless, its sSFR only briefly rises above $\rm 10^{-10} yr^{-1}$ at $z<3$, which does not hinder us from estimating its quiescent visibility factor in epochs at $z>3$.

We calculate the total uncertainty on these reconstructed number densities as
\begin{equation}
 \sigma_{tot}^2 =  \sigma_{N}^2 + \sigma_{poisson}^2 + \sigma_{CV}^2,
\end{equation}
 where $\sigma_{N}$ is the error on the effective number count, $\sigma_{poisson}$ is the Poisson noise term (square root of the effective number count), and $\sigma_{CV}$ is the contribution from cosmic variance.

To estimate $\sigma_{N}$, we combine the uncertainties on the survey completeness and the visibility factor, following the standard error propagation procedure. For the former, we calculate the binomial proportion confidence interval of the completeness fraction $N_{surveyed,\mathbf{m}}/N_{total,\mathbf{m}}$, adopting the Wilson score interval approximation \citep{Wilson.1927}. For the latter, we randomly draw from the inferred SFH parameter posteriors, calculate the $t_{q}'$ for each posterior draw, and convert each $t_{q}'$ to a visibility factor in a given redshift interval, following equation \ref{eq: visibility}. We then estimate asymmetric uncertainties from the 16th and 84th percentile visibility factor posterior draws. If the full posterior corresponds to full quiescent visibility during the given redshift interval (i.e., $\mathcal{P} = 1$), we ignore the uncertainty contribution from the visibility factor. 

The total survey field covered by RUBIES is modest in size ($\rm \sim 300 \,arcmin^2$), and massive galaxies are known to be a highly biased population \citep{Steinhardt.etal.2021, Valentino.etal.2023, Jespersen.etal.2025b}. Therefore, the cosmic variance can dominate the uncertainty budget in the directly measured quiescent number density. To estimate $\sigma_{CV}$, we make use of the cosmic variance code developed in \cite{Jespersen.etal.2025b}, which is based on the method described in \cite{Moster.etal.2010} and utilizes the dark matter halo constraints from the {\sc UniverseMachine} simulations \citep{Behroozi.etal.2019}. We calculate the fractional cosmic variance in a $\rm 0.5 \,dex$ stellar mass bin centered at $\rm M_{*} \sim 10^{11}M_{\odot}$ in three redshift bins of $[2,3]$,$[3,4]$, and $[4,5]$. This stellar mass bin is roughly representative of the stellar mass distribution of quiescent galaxies in this sample. These calculations are performed for EGS and UDS separately by adopting their approximate survey-footprint geometries and then added in inverse quadrature to obtain the total fractional cosmic variance, ranging from $25\% -50\%$ at $z\sim 2-5$. We assume any reconstructed number density at earlier epochs ([$z_{i}$,$z_{i+1}$]) inherits the cosmic variance calculated in the redshift bin where the galaxies are directly observed ([$z_{j}$,$z_{j+1}$]). Finally, we calculate the corresponding reconstructed number densities using SFHs from each given model setup described in Section \ref{sec: Methods}. 

We have also recalculated these reconstructed number densities, adopting $t_{90}$ (time at which 90\% of the stellar mass was formed) instead of $t_{q}$. We find overall similar results, except for the reconstructed number density at $z>5$ from $4<z<5$ populations, where the two model variants with MILES spectral library predict number densities that are $\rm \sim 0.4\,dex$ higher than those calculated with $t_{q}$. The difference in reconstructed number densities computed with either $t_{90}$ or $t_{q}$ reflects the systematic uncertainty in evaluating quiescent visibility or can be physically interpreted as a slower SFR decline in the SFHs of $4<z<5$ quiescent galaxies. We defer a detailed investigation of their SFHs to the future.

\section{Quiescent Galaxy Number Density: Reconstructed Versus Measured}\label{sec: results}
\begin{figure*}[ht!]
\centering
\includegraphics[width = 1\textwidth]{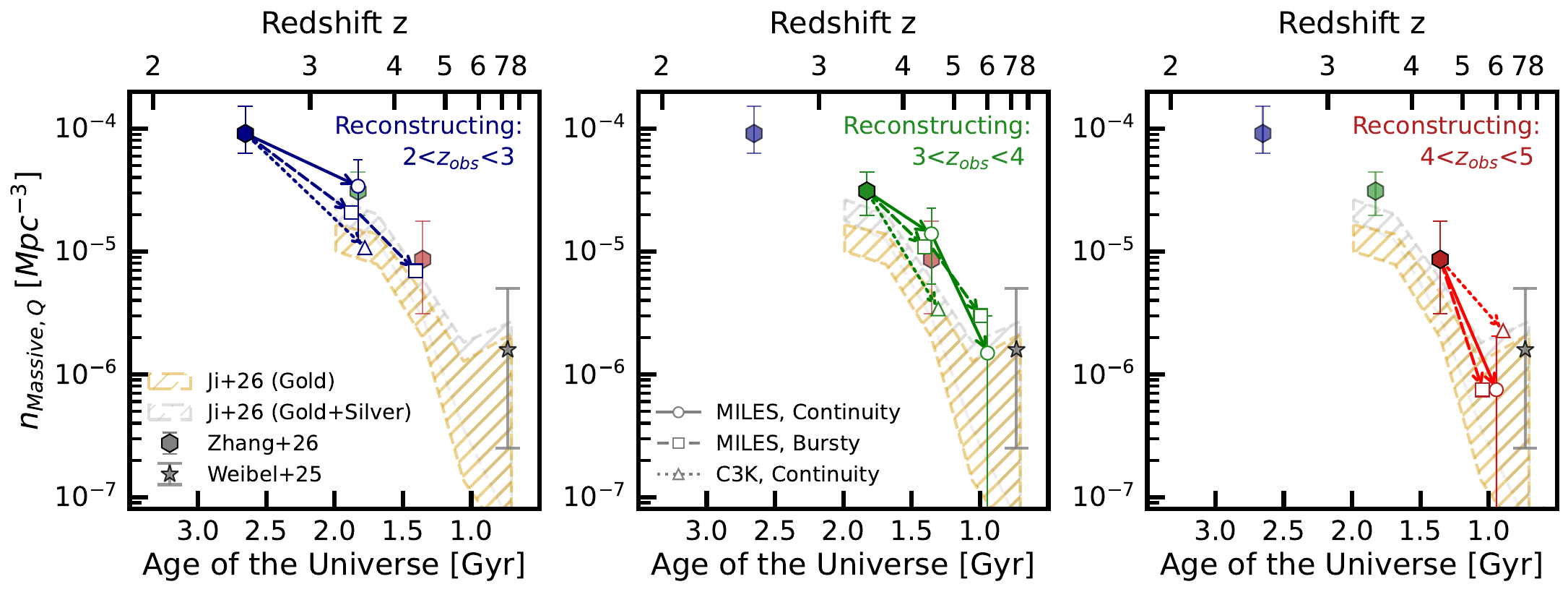}
\caption{In these panels, we compare the number density of quiescent galaxies reconstructed from SFHs in this work (open symbols) to values measured from a subset of direct observations in the literature, including RUBIES (\citealp{Weibel.etal.2025,Zhang.etal.2026}; filled symbols) and PANORAMIC (\citealp{Ji.etal.2026}; hatched bands). We show the number density reconstructed from the $2<z<3$ quiescent population (blue) in the left panel, $3<z<4$ (green) in the middle, and $4<z<5$ (red) in the right. We include uncertainties only on the fiducial model predictions for clarity (open circles). To show the scatter range in these predictions due to modelling assumptions, we show the median model predictions from the bursty prior variant (open squares) and the C3K variant (open triangles). If no quiescent galaxies would have been visible as quiescent in a given redshift bin, we don't show the reconstructed number densities. Overall, the reconstructed quiescent galaxy number densities are in agreement with direct observations within $1 \sigma$ up to $z\sim7$.
\label{fig: punchline}}
\end{figure*}

In Figure \ref{fig: punchline}, we compare the reconstructed quiescent galaxy number densities in this work to those based on direct observations from RUBIES \citep{Weibel.etal.2025, Zhang.etal.2026}. The comparison with any spectroscopic sample of quiescent galaxies to date at $z>5$ is limited by the small sample size, and only a single massive quiescent galaxy has been spectroscopically confirmed at $z\sim7$ by far \citep{Weibel.etal.2025}. Finally, we include quiescent galaxy number densities at $3<z<7$ from \cite{Ji.etal.2026}, which employs the largest quiescent galaxy sample at $z>3$ to date. This comparison sample is collected from over $\rm \sim 1000\,arcmin^2$ JWST $\gtrsim6$-filter NIRCam imaging, consisting of pure parallel imaging from PANORAMIC \citep{Williams.etal.2025} and various premium extragalactic survey fields (e.g., EGS, PRIMER-UDS, PRIMER-COSMOS, GOODS; \citealp{Finkelstein.etal.2023,Bagley.etal.2023,Donnan.etal.2024,Eisenstein.etal.2023a,Eisenstein.etal.2023b}). The number densities uncertainties from PANORAMIC are computed with a high-confidence sub-sample (gold band) and an extended sub-sample (grey band), respectively. Crucially, their associated uncertainties incorporate the field-to-field number density fluctuation, based on the method published in \citep{WeibelAndJespersen.etal.2025}. All studies adopt the same mass limit ($> 10^{10.5} M_{\odot}$), except \cite{Weibel.etal.2025}, which extends slightly lower ($\rm M_{*} \sim 10^{10.2}M_{\odot}$). We note that many other studies have demonstrated this rapidly emerging population \citep[e.g.,][]{Baker.etal.2025b,Yang.etal.2025} and compared growing number densities to theoretical predictions \citep[e.g.,][]{ChandroGomez.etal.2025,Chaikin.etal.2025,Lagos.etal.2025}.

Overall, the reconstructed number densities at any given epoch agree remarkably well (within $1 \sigma$) with higher redshift measurements within RUBIES \citep{Zhang.etal.2026}, especially considering the substantial systematic uncertainties due to the SPS modeling. Furthermore, the overall time evolution also roughly traces the trend observed in the larger PANORAMIC fields \cite{Ji.etal.2026} up to $z\sim7$. 

We also note that the number densities from RUBIES (both the directly measured and the reconstructed from the fiducial model) are slightly higher than PANORAMIC \citep{Ji.etal.2026}, although this is well within the expected RUBIES cosmic variance. This can be in part explained by the intrinsic over-abundance of quiescent galaxies in EGS and UDS at various redshifts. The EGS field is known to contain a proto large-structure (``Cosmic Vine") at $z\sim 3.4$ \citep{Jin.etal.2024} that hosts massive quiescent galaxies \citep[e.g.,][]{Ito.etal.2025}. Similarly, UDS number densities are $\rm \sim 0.2\,dex$ above the average value over multiple sight-lines at $4<z<5$ \citep{Ji.etal.2026}. In addition, it is also possible that the photometric selection of quiescent galaxies employed by PANORAMIC is incomplete at these redshifts. A spectroscopically confirmed quiescent galaxy (``RUBIES-EGS-QG-1" at $z=4.9$; \citealp{deGraaff.etal.2025a}) was classified as ''non-robust" with only photometry \citep{Carnall.etal.2023,Valentino.etal.2023}. 

We emphasize that even given the same data, the inferred progenitor number densities can span a wide range due to modeling choices ($\rm \sim 0.5 \,dex$). We find slightly smaller differences between continuity and bursty SFH priors than \cite{Park.etal.2024}, which is found up to $\rm \sim 1 \,dex$. We suspect that this difference is likely due to two factors. Firstly, we adopt a probabilistic visibility definition (i.e., a galaxy has an effective number count less than one if it was not entirely visible as quiescent throughout the entire redshift bin). Secondly and perhaps more importantly, these systematics likely have less time to dominate at higher redshifts and where galaxies are younger. The quiescent galaxies in RUBIES are predominantly young, with light dominated by A-type stars that typically fade out in $\rm \sim 1 \,Gyrs$ \cite{Zhang.etal.2026}. SFH reconstruction is most precise for that timescale; at older ages, modeling systematics become much more significant. For example, while the bursty model variant predicts some $2<z<3$ RUBIES quiescent galaxies to be still visible as quiescent at $4<z<5$, the other two continuity variants predict no quiescent visibility at all. The quiescent galaxies in \cite{Park.etal.2024} are predominantly older ($>1$ Gyr) and observed at $z\sim2$, thus even recent SFH is more uncertain, and that inference requires extrapolating over a longer time to reach $z>4$.

At $z>7$, all models predict a low or no probability of finding quiescent galaxies with $\rm M_{*}>10^{10.5}M_{\odot}$. While there have been various massive quiescent galaxies with inferred quenching redshifts at $z>7$ in the literature \citep[e.g.,][]{Glazebrook.etal.2024,McConachie.etal.2025a}, no massive ($\rm M_{*}>10^{10.5}M_{\odot}$) quiescent galaxies have been spectroscopically confirmed beyond $z>7$.

Although it is tempting to conclude that the fiducial model among all three best reconstructs the SFHs in these galaxies, since it produces the most self-consistent results between reconstructed and measured number densities in RUBIES. However, we do note that these systematic uncertainties are comparable to the cosmic-variance-dominated scatter in the \cite{Zhang.etal.2026} and \cite{Ji.etal.2026} results. In addition, the sample size of RUBIES is small ($\sim 5$ confirmed objects per redshift bin) and has limited statistical constraining power. Therefore, we cannot yet disfavor any one of these model setups in terms of accurately inferring the SFHs, and we caution against further interpretation.

\section{Discussion and Conclusions}\label{sec: discussion}

We demonstrated that the recent SFHs of quiescent galaxies at $2<z<5$ imply a very consistent population growth with the observed number density evolution since $z\sim7$. This self-consistency solidifies the established tension between the empirical measurements \citep[e.g.,][]{Carnall.etal.2023,Valentino.etal.2023,Baker.etal.2025a,Zhang.etal.2026} and the theoretical models that can produce massive quiescent galaxies in the early Universe \citep[e.g.,][]{Lagos.etal.2025}. This reconstruction is most robust over $\lesssim1$ Gyr timescales and, therefore, is especially effective for younger quiescent galaxies and at high redshift, which helps to mitigate known modeling systematics (age-metallicity, stellar libraries). It is also interesting to extend the reconstruction of SFH beyond quenching and explore the observables (e.g., SED, number density, mass function, luminosity function; \citealp{McConachie.etal.2025a,McConachie.etal.2025b}) of their star-forming progenitors. However, these reconstructions are even more sensitive to additional modeling prior (e.g., dust, SFH parameterization, etc.).

Despite the remarkable agreement between observed and reconstructed number density evolution, there are still a number of limitations to the current study on both sides of the comparison. The most obvious, and easiest to remedy, is the small sample size of this spectroscopic sample, which is further hampered by the limited leverage on cosmic variance \citep{Zhang.etal.2026}.

Additionally, the analysis is sensitive to progenitor bias introduced by, e.g., merging and/or other additions to the population. If most quiescent galaxies at $z<5$ form via several mergers, their progenitors could have dropped out of the selection due to the discrepancy in mass limits, as well as change other fundamental properties (e.g., SFR/color) \citep{Jespersen.etal.2022, Chuang.etal.2024, Cochrane.2025}. However, it is unlikely that major mergers occur frequently enough in the first billion years to become the dominant formation channel. At these redshifts, any major mergers are likely gas-rich and often lead to starbursts that form more stars than they bring in stellar mass \citep{ChandroGomez.etal.2025}. Nevertheless, studies have shown that most early massive quiescent galaxies preferentially reside in over-densities \citep[e.g.,][]{Jespersen.etal.2025,McConachie.etal.2025b}, where merging may be enhanced \citep[e.g.,][]{Huvsko.etal.2023,Shibuya.etal.2025}. Minor mergers also could contribute to population growth at the  $\rm M_{*}>10^{10.5}M_{\odot}$ mass selection limit, since the population is known to extend to lower masses and the mass function is steep \citep{Baker.etal.2025b,Ji.etal.2026}. 

Especially at the highest redshifts, just measuring the number density of massive quiescent galaxies has proven challenging. Thus far, the selection of $z>5$ quiescent galaxies primarily relies on photometry \citep{Baker.etal.2025b,Xiao.etal.2025,Yang.etal.2025,Ji.etal.2026}, with only a handful of spectroscopic confirmations \citep{Onoue.etal.2024,Weibel.etal.2025,Baker.etal.2026}. At lower redshifts ($2<z<5$), dusty star-forming galaxies contamination rates range from $10\%$ to $30\%$ \citep[e.g.,][]{Antwi-Danso.etal.2023,Nanayakkara.etal.2025,Zhang.etal.2026}, but this is unexplored beyond $z\sim5$. Finally, these selections can also be incomplete, which often rely on NIRCam only due to limited MIRI imaging coverage. At $z\sim7$, NIRCam F444W samples the rest-frame optical continuum (i.e., {\it V} band) and template-based extrapolation is required to probe redder rest-frame continuum (i.e., {\it J} band). The lack of sufficient longer wavelength coverage unavoidably inflates the systematic error in redshift estimation as well as quiescence characterization. Although massive quiescent galaxies are bright enough to be detected in NIRCam imaging, \citet{Ji.etal.2026} found that at $z\sim7$ their red colors ($\rm F277W-F444W >1$) can be mistakenly characterized as ``Little Red Dots" \citep[e.g.,][]{Greene.etal.2024,Matthee.etal.2024,Williams.etal.2024}. As existing photometry cannot fully break the degeneracy between these two classes of sources \citep{Hviding.etal.2025}, these objects are often conservatively excluded from number density estimates. If these sources are indeed quiescent galaxies, the corresponding number density at $z\sim 7$ may be inflated up to $\rm n \sim 2\cdot10^{-6} \,Mpc^{-3}$ and become more consistent with extreme ``maximally old" galaxies \citep[e.g.,][]{Glazebrook.etal.2024,deGraaff.etal.2025a,McConachie.etal.2025a}. At $z>5$, it is also possible that a fraction of photometric quasars could otherwise be unidentified quiescent galaxies. The massive quiescent galaxies reported in \cite{Onoue.etal.2024} are dominated by quasar emission in photometry; only spectroscopy reveals the underlying stellar Balmer absorption lines. Ultimately, characterizing the emergence of massive quiescent galaxies -- both directly and via indirect reconstruction from SFHs -- will benefit from probing wider areas with sufficiently deep MIRI coverage and spectroscopic targeting, all of which are achievable with JWST. We advocate for a wider spectroscopic census of quiescent galaxies at $z>4$, in order to map the demographic distribution of extreme quenched timescales, mitigate the systematic uncertainties in photometric selections, and constrain the formation channel of the earliest quiescent galaxies by utilizing the timescale-number density connection.

\begin{acknowledgments}

Some of the data products presented herein were retrieved from the Dawn JWST Archive (DJA). DJA is an initiative of the Cosmic Dawn Center (DAWN), which is funded by the Danish National Research Foundation under grant DNRF140. The Cosmic Dawn Center is funded by the Danish National Research Foundation (DNRF) under grant \#140. 

RB gratefully acknowledges support from the Research Corporation for Scientific Advancement (RCSA) Cottrell Scholar Award ID No: 27587. The work of CCW is supported by NOIRLab, which is managed by the Association of Universities for Research in Astronomy (AURA) under a cooperative agreement with the National Science Foundation. Support for this work was provided by The Brinson Foundation through a Brinson Prize Fellowship grant. 
This work is based in part on observations made with the NASA/ESA/CSA James Webb Space Telescope. The data were obtained from the Mikulski Archive for Space Telescopes at the Space Telescope Science Institute, which is operated by the Association of Universities for Research in Astronomy, Inc., under NASA contract NAS 5-03127 for JWST. These observations are associated with program number 4233. The specific observations analyzed can be accessed via \dataset[doi: 10.17909/sjsj-8p46]{https://doi.org/10.17909/sjsj-8p46}. Support for program no. 4233 was provided by NASA through a grant from the Space Telescope Science Institute, which is operated by the Association of Universities for Research in Astronomy, Inc., under NASA contract NAS 5-03127.

The authors acknowledge the CEERS and PRIMER teams for developing their observing program with a zero-exclusive access period.

\end{acknowledgments}

\facilities{JWST(NIRSpec, NIRCam)}

\software{Astropy \citep{astropy:2013,astropy:2018,astropy:2022}, Numpy \citep{Van.Der.Walt.etal.2011}, Matplotlib \citep{Hunter.2007}, Photutils \citep{photutils}, Prospector\citep{Leja.etal.2017,Johnson.Leja.2017,Johnson.2021}}

\bibliography{sample701}{}
\bibliographystyle{aasjournalv7}

\end{document}